\documentclass[fleqn]{article} 
\pdfoutput=1
\usepackage[bindingoffset=0cm,includeheadfoot,textwidth=17cm,textheight=24.5cm,top=0.5cm]{geometry}
\usepackage{graphicx,algorithmicx,algorithm,algpseudocode,xspace}\usepackage[centertags]{amsmath}
\usepackage{subfigure}

\newcommand{\knotpoint}{knotpoint\xspace}
\newcommand{\tw}{\textwidth}

\newcommand{\knotpoints}{knotpoints\xspace}

\newcommand{\ndist}[3]{{\cal{N}}\br{#1\thinspace\vline\thinspace #2,#3}}
\newcommand{\tp}{t'}

\newcommand{\data}{{\cal{D}}}
\newcommand{\panel}[1]{ \textbf{(#1)}:}

\renewcommand{\subsectionmark}[1]{\markright{}}

\newcommand{\etc}{{\emph{etc.}}}

\newcommand{\bigo}[1]{O\br{#1}}

\newcommand{\br}[1]{\left( {#1} \right)}
\newcommand{\sq}[1]{\left[ {#1} \right]}

\newcommand{\boldmu}{\boldsymbol{\mu}}

\newcommand{\boldSigma}{\boldsymbol{\Sigma}}

\newcommand{\pgp}{p_{\tiny{GP}}}
\newcommand{\pode}{p_{\tiny{ODE}}}

\renewcommand{\v}[1]{\mathbf{#1}}

\newcommand{\secref}[1]{section(\ref{#1})}
\renewcommand{\algref}[1]{algorithm(\ref{#1})}
\newcommand{\figref}[1]{figure(\ref{#1})}

\newcommand{\beq}{\begin{equation}}
\newcommand{\eeq}{\end{equation}}

\newcommand{\xdot}{\dot{x}}
\newcommand{\xdotdot}{\ddot{x}}

\newcommand{\vxdot}{\dot{\v{x}}}

\newcommand{\tdiff}[2]{\frac{d}{d{#2}}{#1}}
\newcommand{\pdiff}[2]{\frac{\partial}{\partial{#2}}{#1}}
\newcommand{\pdifff}[3]{\frac{\partial^2}{\partial{#2}\partial{#3}}{#1}}

\newcommand{\tdiffs}[2]{\frac{d^2}{d{#2}^2}{#1}}
\newcommand{\ax}{\tilde{x}}

\newcommand{\ocm}{\hspace{1cm}}
\newcommand{\hcm}{\hspace{0.25cm}}
\newcommand{\av}[1]{\left\langle{#1}\right\rangle}
\newcommand{\trans}{^{\textsf{T}}}
\title{On solving Ordinary Differential Equations using Gaussian Processes}
\author{David Barber\\Department of Computer Science\\University College London}

\newcommand{\xt}{x_{n}}
\newcommand{\xtp}{x_{n+1}}
\newcommand{\axtp}{\tilde{x}_{n+1}}
\newcommand{\xdott}{\xdot_{n}}

\begin{document}
\maketitle

\begin{abstract}
We describe a set of Gaussian Process based approaches that can be used to solve non-linear Ordinary Differential Equations. We suggest an explicit probabilistic solver and two implicit methods, one analogous to Picard iteration and the other to gradient matching.  All methods have greater accuracy than previously suggested Gaussian Process approaches. We also suggest a general approach that can yield error estimates from any standard ODE solver.
\end{abstract}

\section{The Initial Value Problem}

Given an Ordinary Differential Equation (ODE) with known initial condition $x(t_1)=x_1$
\beq
\tdiff{x(t)}{t}=f(t,x(t),\theta)
\eeq
the Initial Value Problem (IVP) is to find the differentiable function $x(t)$ over some specified time interval $t\in\sq{t_1,t_T}$ that satisfies the ODE subject to the initial value condition. In general $x(t)$ is a vector so that higher order scalar ODEs can be embedded as first order vector ODEs \cite{Robinson2004}. In general this problem requires an approximate numerical solution and we denote the approximation  at time $t_n$ to $x(t_n)$ by $x_n$. There is a vast literature on this topic (see \cite{Robinson2004} for an introduction) and several families of techniques that can be applied  such as one-step methods, multistep methods, fixed and variable step length, and implicit and explicit approaches. In general there is no single `best' method with the methods having different properties in terms of numerical stability, speed, number of function $f$ evaluations, parallelisabiltiy \etc 

Recently there has been interest in the machine learning community in the application of Gaussian Processes for the IVP\footnote{The more general boundary value problems can also be addressed using related approaches.} \cite{Skilling91,HennigHauberg14,Chkrebtii2013} and estimation of ODE parameters given potentially noisy observations $\data$ \cite{Calderhead2008,Dondelinger2013,WangBarber14}. An ideal approach to parameter estimation is based on Bayesian Numerical Integration. Writing $t_1,\ldots, t_N$ for the times at which data is observed and $x(t_n)$ for the true solution to the IVP at those times, 
\beq
p(\theta,x_{2:N}|x_1,\data)\propto p(\data|x_{2:N})p(x_{2:N}|x_1,\theta)
\eeq
where the term $p(x_{2:N}|x_1,\theta)$ represents a distribution over true solutions given the initial value. Generally this otherwise ideal approach is problematic since classical IVP solution techniques do not produce a distribution over solutions $x_{2:N}$, meaning that the uncertainty (which must exist due to the numerical approximation) in the solution is not correctly accounted for. 

The approaches in \cite{Calderhead2008,Dondelinger2013,WangBarber14} use Gaussian Processes to circumvent the requirement to produce $p(x_{2:N}|x_1,\theta)$ and work by implicitly fitting an alternative function to the data whose gradient must match the gradient specified by the ODE at the observation times.
Whilst these recent parameter estimation approaches that avoid the requirement to find $p(x_{2:N}|x_1,\theta)$ look promising, it nevertheless remains of interest to find distributions $p(x_{2:N}|x_1,\theta)$ since these can be used to solve the IVP and characterise uncertainty in the solution. This is the focus of this work, in which we assume that the parameters $\theta$ of the ODE are known, but an estimate of uncertainty in the solution is required\footnote{We therefore assume $\theta$ is fixed and known and drop the notational conditioning on $\theta$ throughout.}. 

\subsection{Gaussian Processes and Linear ODEs}

In the case that $f$ is linear in $x(t)$, the solution involves integrals of matrix exponentials, which generally cannot be computed in closed form but can be approximated using for example the Magnus expansion \cite{Robinson2004}.
An alternative approximate approach that avoids explicit integration and generalises the solution to the case of additive Gaussian noise is to assume that $x(t)$ follows a Gaussian Process (GP) with covariance function $C(t,\tp)$, see for example \cite{Graepel-03}. Then writing $f(t,x(t))$ in terms of a matrix $L$, time-varying term, $\phi(t)$ and Gaussian noise $\epsilon(t)$
\[
f(t,x(t)) = L x(t) - \phi(t) + \epsilon(t)
\]
we have
\beq
y(t)\equiv  \xdot(t) -  L x(t) - \epsilon(t) = \phi(t), \ocm\text{where }\hcm
\xdot(t)\equiv \tdiff{x(t)}{t}
\eeq
Since $x(t)$ is assumed a GP, and $y(t)$ is a linear function of $x(t)$ and $\epsilon(t)$, then $y(t)$ is also a GP. The covariance function of this new process is straightforward to obtain using the standard rules, see  \cite{Rasmussen2006}. For example, the covariance terms involving $\xdot$ are simply obtained by differentiating the covariance function of $x$:
\beq
\av{\xdot(t)x(\tp)}_{p(\xdot,x)}=\pdiff{C(t,\tp)}{t}, \ocm \av{\xdot(t)\xdot(\tp)}_{p(\xdot)}=\pdifff{C(t,\tp)}{t}{\tp}
\eeq
where $\av{f(x)}_{p(x)}$ denotes expectation of the function $f(x)$ with respect to the distribution $p(x)$. Given then observations $y_n\equiv\phi(t_n)$ at the given observation times and any boundary or initial conditions on $x$ and $\xdot$, then $x(t)$ is a GP whose mean and covariance function is given by the standard Gaussian conditioning formulae
\beq
p(x|y)=\ndist{x}{\mu_x+C_{xy}C_{yy}^{-1}(y-\mu_y)}{C_{xx}-C_{xy}C_{yy}^{-1}C_{yx}}
\eeq
In this way we can globally approximately solve the IVP (or BVP), giving a Gaussian distribution over the solution approximation $p(x_{2:N}|x_1,\data)$. Whilst this method has cubic complexity in the number of points $N$ that we need to evaluate the function at, this will typically be much smaller than the number of timepoints used in a standard ODE solver, provided that the solution is sufficiently smooth.

%

\subsection{Skilling's IVP approach for non-linear ODEs}

In the case that $f$ is not linear in $x$, the problem is generally much more complex. One approach would be to assume that the approximation follows a GP and perform local linearisation, analogous to Exponential Integrators, see for example \cite{hochbruck2010}. However, recent work in this area \cite{HennigHauberg14,Chkrebtii2013} has developed the suggestion by Skilling \cite{Skilling91}, which we outline below. 



\newcommand{\tset}{{\cal T}}
\newcommand{\kset}{{\cal K}}
\newcommand{\iset}{{\cal I}}

The fundamental quantity of interest in Skilling's \cite{Skilling91} approach\footnote{It is perhaps worth mentioning that Skilling viewed his approach as only a suggestion amongst other potential related approaches.} for the IVP is the set of derivatives  $\vxdot\equiv \xdot_1,\ldots,\xdot_N$  at specified `\knotpoints' $t_1,\ldots,t_N$. In \cite{HennigHauberg14,Chkrebtii2013} a GP is assumed for the approximate solution $x_{1:N}$. 
We start with the known initial state $x_1$ and compute its derivative\footnote{We drop the potential dependence of $f$ on $t$ to avoid notational clutter.} $\xdot_1=f(x_1)$.  This is the only point and derivative that we know with certainty. One can interpret this as an observation of the derivative with zero observation error, $\sigma^2_1=0$. We assume a zero mean GP with known covariance function\footnote{The approach in \cite{Chkrebtii2013} is slightly different -- they do not condition on knowing $x_1,\xdot_1$ (see their equation 11). Rather they impose the mean of the GP at timestep 1 to be $x_1,\xdot_1$ with zero covariance. Subsequent timesteps have zero mean. We take the approach as outlined in \cite{HennigHauberg14} -- there is little practical difference in the two approaches.}. Using the GP we can form a distribution for the solution at the next \knotpoint
\beq
\pgp(x_2|x_1,\xdot_1)
\label{eq:sk:post}
\eeq
We now sample a value for $x_2$ from \eqref{eq:sk:post} and subsequently compute the derivative $\xdot_2=f(x_2)$. Note that both the value $x_2$ and derivative $\xdot_2$ will not necessarily correspond to the true solution and its derivative. Because of this, only the derivative is retained and 
the interpretation is that one has observed the derivative $\xdot_2$ with measurement error $\sigma^2_2$ specified by the variance of $\pgp(\xdot_2|x_1,\xdot_1)$.  Given $x_1,\xdot_1=f(x_1),\xdot_2=f(x_2)$ and the corresponding variances on these observations $\sigma^2_{1:2}$, one now forms the GP prediction
\beq
\pgp(x_3|,x_1,\xdot_1=f(x_1),\xdot_2=f(x_2),\sigma^2_{1:2})
\eeq
As before one then samples a value for $x_3$ and subsequently computes the derivative observation $\xdot_3=f(x_3)$ which is assumed to be measured with observation variance obtained from $\pgp(\xdot_3|\xdot_{1:2},\sigma^2_{1:2}
)$. One continues in this manner defining a set of derivative observations $\xdot_{1:N}$ and corresponding observation noises $\sigma^2_{1:N}$. These can then be used as part of a standard GP prediction model to infer $\pgp(x_{2:N}|x_1,\xdot_{1:n},\sigma^2_{1:n})$.

\begin{figure}[t]
\begin{center}
\subfigure[$x_1(t)$]{\includegraphics[width=0.49\tw]{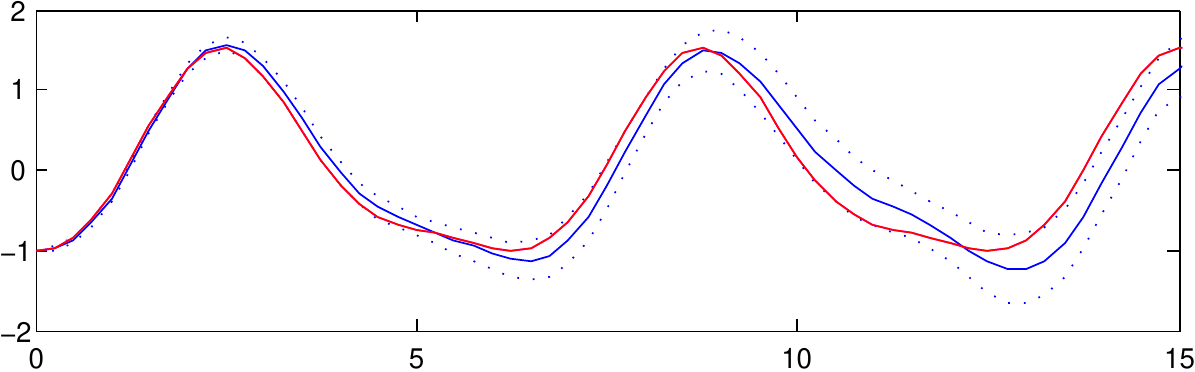}}\hcm
\subfigure[$x_2(t)$]{\includegraphics[width=0.49\tw]{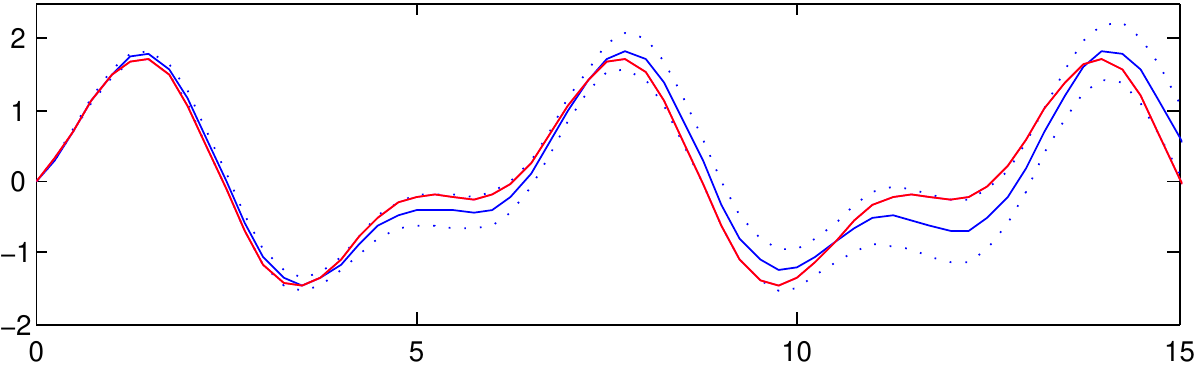}}
\end{center}
\caption{Solutions to the ODE \eqref{eq:simple:ode} for  $\theta=2$. Plotted are the solution (left) and its derivative (right). The exact solution is plotted in red. Two solution methods are shown: Runge-Kutta4.5 (magenta), which is virtually indistinguishable from the exact solution, and the Skilling GP approach with  stepsize $\delta=0.25$. Plotted in dashed lines are the estimated one standard deviation errors in the GP solution.  
 \label{fig:simple:skilling}}
\end{figure}

Whilst this procedure can be shown to retrieve the exact integrated curve in the limit of infinitely densely spaced \knotpoints \cite{Chkrebtii2013}, the naive time complexity is $\bigo{N^3}$ (due to Gaussian conditioning) which would most likely make this much slower than standard ODE solvers.  This complexity can however be reduced by using more specialised covariance functions, see \cite{Chkrebtii2013}.
In \figref{fig:simple:skilling} we show this approach applied to solving the ODE\footnote{Note that here the subscript denotes the component of the two dimensional vector exact solution.}
\beq
f_1(t) = x_2(t), \ocm f_2(t) = -x_1(t)+\sin(\theta t);
\label{eq:simple:ode}
\eeq
which has exact solution
\[
x_1(t)=\br{-\theta^2\cos(t) + \theta\sin(t) - \sin(\theta t) + \cos(t)}/(\theta^2-1), \hcm x_2(t)=dx_1(t)/dt
\]
For this experiment (and throughout the paper) we used the squared exponential covariance $C(t,\tp)=\exp(-(t-\tp)^2)$. From \figref{fig:simple:skilling} we see that the Skilling GP procedure is substantially worse in terms of numerical accuracy than the standard Runge Kutta approach. One potential reason for this is that it discards useful information gathered about the function, namely the sample values $x_2,\ldots,x_N$. These could also be included as `noisy' measurements of the true integrated curve, with measurement error similarly given by the variance of the predicted GP. However, our experience is that extending the scheme in this manner does not significantly improve the accuracy of the approach.  Given the drawbacks of this GP solution technique, we were motivated to consider alternative approaches for probabilistic solutions and uncertainty estimates in non-linear ODEs.

\section{Novel ODE solvers}

There are a great many directions that one could take in constructing a probabilistic solver and we outline only three. We also describe in \secref{sec:error:est} a general method that can be used to estimate the error in any ODE solution (obtained from a standard ODE solver).  

\begin{figure}[t]
\begin{center}
\subfigure[$x_1(t)$]{\includegraphics[width=0.49\tw]{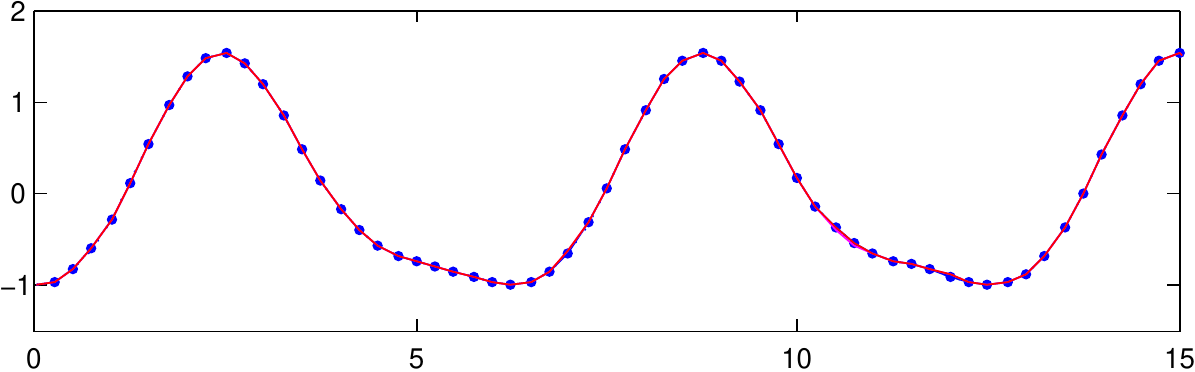}}\hcm
\subfigure[$x_2(t)$]{\includegraphics[width=0.49\tw]{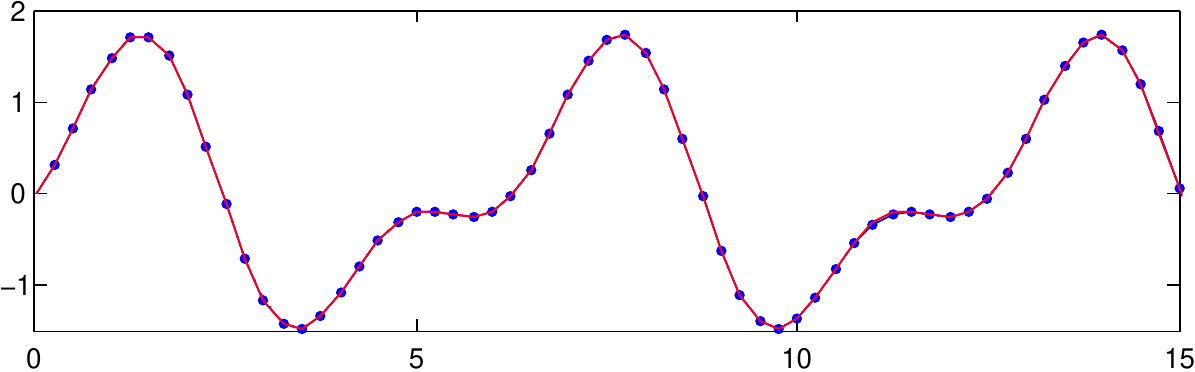}}
\end{center}
\caption{Solutions to the ODE \eqref{eq:simple:ode} for  $\theta=2$. Plotted are the solution (left) and its derivative (right). The exact solution is plotted in red and the Explicit GP approach with  stepsize $\delta=0.25$ is plotted in blue dots (indistinguishable from the the exact solution). The error estimates are too small to be visible.
 \label{fig:simple:explicit}}
\end{figure}

\subsection{An Explicit Solver}
Our first solver follows most closely in spirit to Skilling's approach. Given $x_1,\xdot_1$, we would like to find $p(x_2|x_1,\xdot_1)$. We can obtain this using\footnote{We write out the steps explicitly to explain the intuition behind the derivation.}
\begin{align*}
p(x_2|x_1,\xdot_1)&=\int_{\xdot_2} p(x_2,\xdot_2|x_1,\xdot_1)=\int_{\xdot_2} \pgp(x_2|\xdot_2,x_1,\xdot_1)p(\xdot_2|x_1,\xdot_1)\\
&=\int_{\xdot_2} \pgp(x_2|\xdot_2,x_1,\xdot_1)\int_{\ax_2}p(\xdot_2,\ax_2|x_1,\xdot_1)\\
&=\int_{\xdot_2} \pgp(x_2|\xdot_2,x_1,\xdot_1)\int_{\ax_2}\pode(\xdot_2|\ax_2,x_1,\xdot_1)\pgp(\ax_2|x_1,\xdot_1)\\
&=\int_{\xdot_2,\ax_2} \pgp(x_2|\xdot_2,x_1,\xdot_1)\pode(\xdot_2|\ax_2,x_1,\xdot_1)\pgp(\ax_2|x_1,\xdot_1)
\end{align*}
For simplicity we assume a deterministic ODE so that we can write
\begin{align*}
p(x_2|x_1,\xdot_1)&=\int_{\xdot_2,\ax_2} \pgp(x_2|\xdot_2,x_1,\xdot_1)\delta(\xdot_2-f(\ax_2))\pgp(\ax_2|x_1,\xdot_1)\\
&=\int_{\ax_2} \pgp(x_2|\xdot_2=f(\ax_2),x_1,\xdot_1)\pgp(\ax_2|x_1,\xdot_1)
\end{align*}
We can then obtain samples from $p(x_2|x_1,\xdot_1)$ by forward sampling: we sample a putative future value of the solution using the GP $\pgp(\ax_2|x_1,\xdot_1)$, conditioned on past information. However, this value does not necessarily satisfy the derivative requirement of the ODE. To see how well it matches, we calculate what the derivative $\xdot_2$ of this putative solution should be. Given this we can then sample a value for $x_2$. In this manner the generated $x_2$ will be consistent with the smoothness assumption of the GP and also consistent with the derivative requirement of the ODE\footnote{Although we do not do so here, one can  consider variations on this theme such as including additional putative values such as $\tilde{\tilde{x}}_2$ \etc}. 

%

More generally, given a sample from the distribution 
\[
p(\xt|x_{1:n-1},\xdot_1)
\]
the distribution $p(\xtp|x_{1:n},\xdot_1)$ is recursively defined by, see \algref{alg:explicit},
\begin{multline*}
\hspace{-0.75cm}p(\xtp|x_{1:n},\xdot_1)\\
\hspace{-0.75cm}=\int_{\axtp}\pgp(\xtp|x_{1:n},\xdot_{1:n}=f(x_{1:n}),\xdot_{n+1}=f(\axtp))\pgp(\axtp|x_{1:n},\xdot_{1:n}=f(x_{1:n}))
\end{multline*}
Given $x_{1:n}$ and $\xdot_{1:n}$, we then sample a state $\axtp$ from $\pgp(\axtp|x_{1:n},\xdot_{1:n})$ and subsequently a state $\xtp$ from 
$\pgp(\xtp|,x_{1:n},\xdot_{1:n}=f(x_{1:n}),\xdot_{n+1}=f(\axtp))$. We repeat this process until time index $N$, which defines then a single trajectory $x_{2:N}$. To define another solution sample $x_{2:N}$, we repeat the above process beginning from time index $n=1$. The distribution over solutions is then formally obtained by $\prod_n p(\xt|x_{1:n-1},\xdot_1)$.
%

Note that this procedure differs significantly from Skilling's. Firstly, points are generated that are more likely to be consistent with the ODE requirement. Also, by conditioning on the past samples, we can limit the time horizon for the GP prediction. That is, at timestep $n$, rather than conditioning on all past observations $x_{1:n-1}$ (which would have computational complexity cubic in $n$) we can limit the conditioning to say $M$ previous observations, limiting the complexity of drawing a sample for $\xt$ to cubic complexity in $M$. This is a significant improvement in complexity than previous approaches and brings the method in line with standard multistep ODE solver complexities. 

We demonstrate the method in \figref{fig:simple:explicit} which has the same setup as Skilling's approach in \figref{fig:simple:skilling}. Despite using the same stepsize $\delta=0.25$ in both approaches,  the explicit GP method has excellent comparative performance and is  computationally significantly cheaper.  

\algsetlanguage{pseudocode}
\begin{algorithm}[t]
\caption{Explicit Multistep ODE solver \label{alg:explicit}. Draw $S$ samples, each with an $M$ length history.}
\begin{algorithmic}[0]
\For {$l=1:S$} \Comment{Sample multiple trajectories}
\For {$n=1:N$} \Comment{Sample a single trajectory}
\State Draw a sample $\axtp$ from $\pgp(\axtp|x_{n-M:n},\xdot_{n-M:n})$
\State Compute the derivative at this point $\xdot_{n+1}=f(\axtp))$
\State Draw $\xtp$ from 
$\pgp(\xtp|,x_{n-M:n},\xdot_{n-M:n}=f(x_{n-M:n}),\xdot_{n+1}=f(\axtp))$. 
\EndFor
\State This defines a sample $x^l_{1:N}$
\EndFor
\end{algorithmic}
\end{algorithm}

\algsetlanguage{pseudocode}
\begin{algorithm}[t]
\caption{Implicit Multistep ODE solver \label{alg:implicit:sample}. Draw a sample solution}
\begin{algorithmic}[0]
\State Initialise the sample $x^1_{2:N}$
\For {$i=1:I$} \Comment{Iteration counter}
\For {$n=2:N$} \Comment{Sample a trajectory}
\State Compute the derivative at each point $\xdot_{n}=f(x^i_n)$
\EndFor
\State Draw a sample $x_{2:N}$ from 
$\pgp(x_{2:N}|\xdot_{1:n},x_1)$. 
\EndFor
\State After $I$ iterations we have a sample $x_{1:N}$
\end{algorithmic}
\end{algorithm}

\subsection{An Implicit Solver}

A drawback of explicit approaches is that they can lack consistency and also stability \cite{Robinson2004}. One way to view deriving consistent approximations is to require that if we solve going forwards, and then using this solution reverse time and solve backwards, we should end up where we started from\footnote{This is the intuition behind for example the mid-point extension of the Euler method.}. A related approach is to assume a solution that must be globally consistent\footnote{This is essentially the approach taken by Picard iteration.}. We will first assume that we wish to sample a trajectory $p(x_2,x_3|x_1,\xdot_1)$. We can write (for a deterministic ODE) 
\begin{align*}
\hspace{-1cm}p(x_2,x_3|x_1,\xdot_1)&=\int_{\xdot_2,\xdot_3}p(x_2,x_3,\xdot_2,\xdot_3|x_1,\xdot_1)\\
&=\int_{\xdot_2,\xdot_3}p(x_2,x_3|x_1,\xdot_{1:3},x_1)p(\xdot_2,\xdot_3|x_1,\xdot_1)\\
&=\int_{\ax_2,\ax_3,\xdot_2,\xdot_3}\pgp(x_2,x_3|x_1,\xdot_{1:3})\pode(\xdot_2,\xdot_3|\ax_2,\ax_3,x_1,\xdot_1)p(\ax_2,\ax_3|x_1,\xdot_1)\\
&=\int_{\ax_2,\ax_3}\pgp(x_2,x_3|\xdot_1,\xdot_2=f(\ax_2),\xdot_3=f(\ax_3),x_1)p(\ax_2,\ax_3|x_1,\xdot_1)
\end{align*}
Thus, if we start with a distribution $p(\ax_2,\ax_3|x_1,\xdot_1)$ over solutions, the above updates this to a new distribution $p(x_2,x_3|x_1,\xdot_1)$. This is analogous to Picard iteration, see for example \cite{Robinson2004}, and can be perhaps best considered as a distributional approximation to the  Picard approach. By recursing, we seek the fixed point solution $p^*(x_2,x_3|x_1,\xdot_1)$ of the above procedure. The fixed point then has the required global consistency property\footnote{Intuitively, in the limit of small $\delta$ this tends to the Picard iteration and thus to the exact solution, due to the contraction property of the Picard operator.}. 

Since the above updating process is not closed with respect to any standard distribution class, one could alternatively draw samples recursively, see \algref{alg:implicit:sample}. Another approach, which we adopt in the experiments, is to assume that $p(\ax_2,\ax_3|x_1,\xdot_1)$ is approximated by a Gaussian with mean $\boldmu$ and covariance $\boldSigma$. We then evaluate 
\[
\pgp(x_2,x_3|\xdot_1,\xdot_2=f(\mu_2),\xdot_3=f(\mu_3),x_1)
\]
which defines the new mean and covariance. We iterate this to convergence.
%
%

In principle, one can apply the above to the whole solution trajectory. However, this is computationally expensive, scaling $\bigo{N^3}$ due to the Gaussian conditioning step. An alternative is to consider a small size $M$ window and solve for the future values in this small window. Then the window is moved forward one timestep. For our example above, this would define a distribution for $x_2$ and $x_3$. We could then move forward one timestep with $x_2$ replacing $x_1$ as the conditioned information. Similarly, rather than $x_1$ being observed with certainty, we assume that $x_2$ is observed with variance obtained from $p^*(x_2|x_1,\xdot_1)$. As we move forwards, we can retain a limited history of the computed values to bracket the variable $\xt$ by a small number of past and future variables, analogous to implicit multistep solvers \cite{Robinson2004}.

%
%
%
%
%

\begin{figure}[t]
\begin{center}
\subfigure[$x_1(t)$]{\includegraphics[width=0.49\tw]{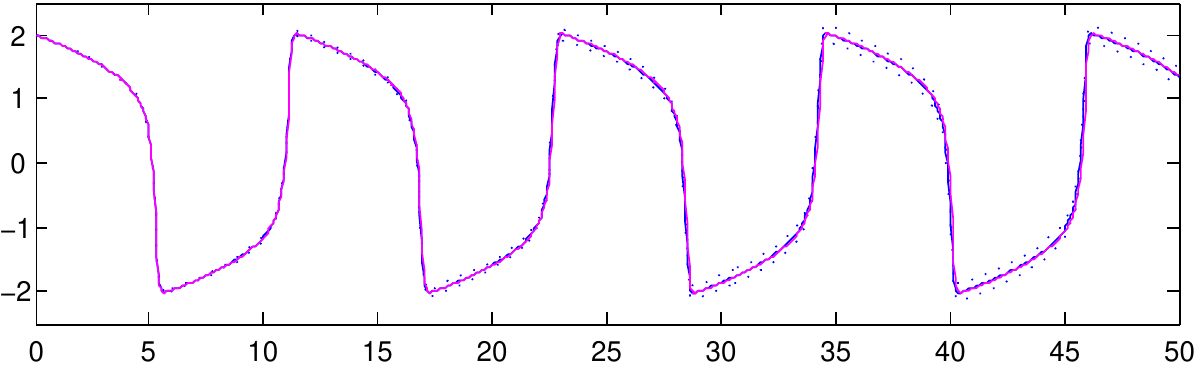}}\hcm
\subfigure[$x_2(t)$]{\includegraphics[width=0.49\tw]{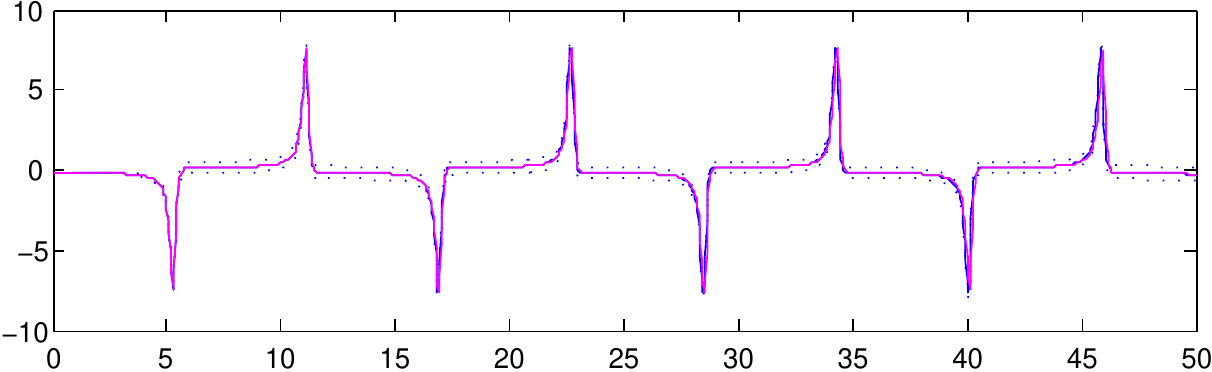}}
\end{center}
\caption{Van der Pol oscillator \eqref{eq:vdp} with $\theta=5$. Plotted are the solution (left) and its derivative (right). Two solution methods are shown: Runge-Kutta4.5 (magenta) and the Implicit GP approach with window length 5 and stepsize $\delta=0.05$. Plotted in dashed lines are the estimated errors in the GP solution.  
 \label{fig:vdp:mid}}
\end{figure}

As an example we show in \figref{fig:vdp:mid} the solution technique applied to the Van der Pol oscillator.
\beq
f_1(x(t))=x_2(t), \ocm f_2(x(t))=-x_1(t)+\theta\br{1-x_1^2(t)}x_2(t)
\label{eq:vdp}
\eeq
As we can see, the approach performs well, with increasing uncertainty  as time increases. Compared to the Explicit GP approach (not shown) the Implicit approach solves this problem more accurately, though with a larger number of function evaluations due to the fixed point iteration. 

\subsection{Implicit Gradient Matching}

\begin{figure}[t]
\begin{center}
\subfigure[$x_1(t)$]{\includegraphics[width=0.49\tw]{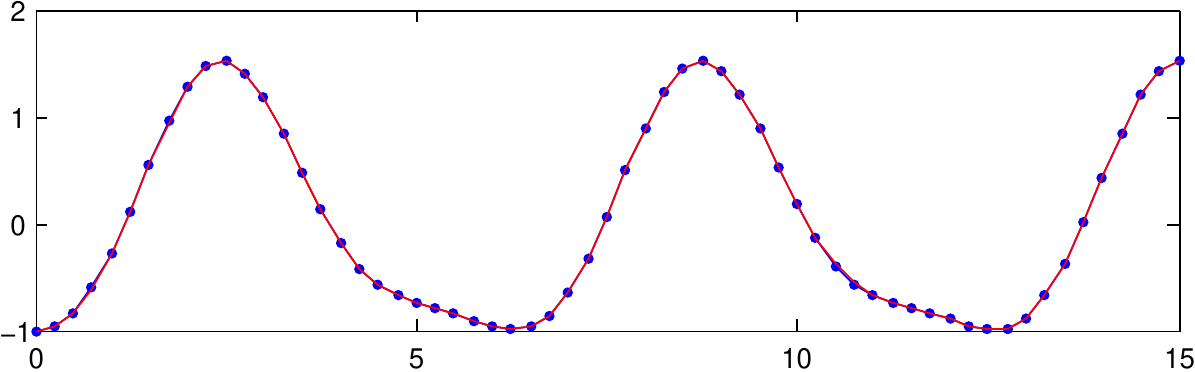}}\hcm
\subfigure[$x_2(t)$]{\includegraphics[width=0.49\tw]{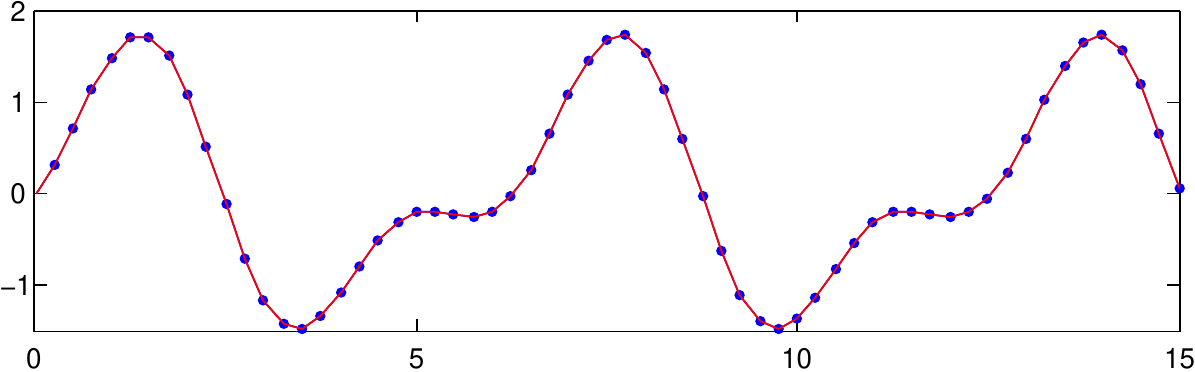}}
\end{center}
\caption{Solutions to the ODE \eqref{eq:simple:ode} for  $\theta=2$. Plotted are the solution (left) and its derivative (right). The exact solution is plotted in red. The gradient matching approach with window of length 10 is shown with stepsize $\delta=0.25$ and is virtually indistinguishable from the exact solution. 
 \label{fig:simple:grad}}
\end{figure}

If we assume we are given a proposed solution $x_{1:N}$, we can use a GP to calculate the derivative distribution of this point sample curve,
\[
\pgp(\xdot_{2:N}|\xdot_1,x_{1:N})
\]
A self consistency requirement is that these derivatives should match the known ODE derivatives $f(\xt)$ at the points $t=2,\ldots, N$. The expected mismatch is
\[
E(x_{2:T})\equiv \av{\sum_{\tau=2}^N \br{f(x_\tau)-\xdot_\tau}^2}_{\pgp(\xdot_{2:N}|\xdot_1,x_{1:N})}
=\sum_{\tau=2}^N \br{f(x_\tau)-\av{\xdot_\tau|\xdot_1,x_{1:N}}}^2+\sigma^2(\xdot_\tau)
\]
The term $\av{\xdot_\tau|\xdot_1,x_{1:N}}$ denotes the mean of the variable $\xdot_\tau$ conditioned on knowing $\xdot_1,x_{1:N}$. For a GP, the final variance term $\sigma^2(\xdot_\tau)$ is independent of $x_{2:N}$. Also for a GP, the predicted mean is a linear function of the observation and thus the mean of $\xdot_\tau$ is a linear function of $\xdot_1,x_{1:n}$. An equivalent optimisation problem is to minimise with respect to $\v{x}$
\[
F(\v{x})\equiv \sum_{\tau=2}^N \br{f(x_\tau)-c_\tau -\v{a}_\tau\trans\v{x}}^2
\]
where $\v{x}=x_{2:N}$ for suitably defined vectors $\v{a}_\tau$ and constants $c_\tau$ (these are simply derived from the GP). The optimisation can be achieved by standard approaches. In our experiments, we formed an update based on equating the derivative of $F$ to zero; this gives a rapidly converging estimate.  Whilst one can in principle carry out this optimisation for all $x_{2:N}$, this is wasteful since only timepoints close to the initial time will be relevant for determining the solution close to the initial time (the solution method will first determine $x_2$ and then $x_3$, \etc).  For this reason we therefore considered a windowed approach, moving the solution forward by one timestep once a convergence criterion for the window is passed.  An example is given in \figref{fig:simple:grad}. In our experience, this gradient matching approach performs well, but is less accurate than the implicit GP approach.

\subsection{Using Higher Order Derivative Information}
One benefit of the GP approach is that it is straightforward to extend to conditioning on higher order derivatives. Given the collection of ODEs, 
\[
\tdiff{x_i}{t}=f_i(t,x)
\]
we can compute 
\[
\tdiffs{x_i}{t}=\pdiff{f_i(t,x)}{t}+\sum_j\tdiff{f_i(t,x)}{x_j}\tdiff{x_j}{t}
=\pdiff{f_i(t,x)}{t}+\sum_j J_{ij}(t,x)f_j(t,x)
\]
where the Jacobian is defined
\[
J_{ij}(t,x)\equiv \tdiff{f_i(t,x)}{x_j}
\]
We can then use these second order derivatives $\xdotdot$ as part of the GP conditioning set. Our code includes the option of using higher order derivative information in any of the above three novel solvers.  

\subsection{Deriving Error Estimates\label{sec:error:est}}

Given an approximate solution $x_{2:N}$ for the IVP from a standard ODE solver, we can estimate the error as follows. We first compute $\pgp(\xdot_{2:N}|\xdot_1,x_{1:N})$ and draw a sample $\xdot_{2:N}$ from this. If our solution $x_{2:N}$ were correct, then the derivative should be $f(x_{2:N})$. We can therefore obtain a local estimate of the error in the derivative $\xdott$ by
\beq
\dot{\sigma}^2_n\equiv \av{ \br{f(\xt) - \xdott}^2}_{\pgp(\xdott|\xdot_1,x_{1:N})}
\label{eq:xdot:var}
\eeq
Using this we can then form the Gaussian likelihood for an ODE solution $\pgp(x_{2:N}|x_1,\xdot_1,\xdot_{2:N}=f(x_{2:N}))$ in which during the GP conditioning it is assumed that the derivatives are observed with the variances computed by \eqref{eq:xdot:var}. This likelihood can be used to assess the quality of the ODE solution $x_{2:N}$.

\section{Discussion and Summary}

There has been recent interest in approximate methods for solving ODEs based on using Gaussian Processes. We have noted that the approaches \cite{Chkrebtii2013,HennigHauberg14} based on Skilling's suggestion \cite{Skilling91} suffer some drawbacks and it is unclear if they can be made practically useful in their current form.  In contrast, we suggested a collection of techniques based on insights from standard ODE solvers, using both implicit and explicit information.  To date we have carried out only limited experiments but believe these are promising directions to consider as alternatives to existing GP approaches for solving ODEs. 

The simplest of our approaches is the Explicit GP method which has reasonable accuracy and is analogous to Explicit multistep ODE solvers. In our experience, as would be expected from such a simple forward explicit approach, the accuracy is lower (for a similar number of function evaluations) than can be achieved by more sophisticated implicit techniques. 

Our Implicit GP method is also straightforward to implement, though is slightly more complex than the forward approach. However, the numerical accuracy of the approach is high. In our experiments on the Van der Pol oscillator, the method outperforms the Explicit approach and has accuracy similar to standard ODE solvers such as Runge Kutta 4.5.

The gradient matching approach is another implicit approach that also improves on the Explicit GP approach. In our experiments, we have found that the method has comparable accuracy to the Implicit approach though solving the required optimisation problem at each timestep is more costly than the fixed-point iteration of the Implicit GP approach.

The extension of these methods to solving partial differential equations, as in \cite{Chkrebtii2013}, is in principle straightforward.  

\subsubsection*{Acknowledgements}

I would like to thank Mark Girolami for helpful discussions.



\end{document}